\documentclass[twocolumn,preprintnumbers,superscriptaddress,nofootinbib,aps,prd,floatfix]{revtex4}

\usepackage{enumerate}
\usepackage{amsmath,amssymb}
\usepackage{graphicx}
\usepackage{bbm,slashed}
\usepackage{bbold}
\usepackage{xspace,slashed}
\usepackage{hyperref}
\hypersetup{colorlinks=true, citecolor=blue, urlcolor=blue, linkcolor=blue}
\usepackage[normalem]{ulem}
\usepackage{subfigure}
\usepackage{multirow,array}

\usepackage{tabularx,booktabs} %midrule toprule etc in tables
\newcolumntype{C}{>{$}c<{$}}
\AtBeginDocument{%
  \heavyrulewidth=.08em
  \lightrulewidth=.05em
  \cmidrulewidth=.03em
  \belowrulesep=.65ex
  \belowbottomsep=0pt
  \aboverulesep=.4ex
  \abovetopsep=0pt
  \cmidrulesep=\doublerulesep
  \cmidrulekern=.5em
  \defaultaddspace=.5em
}
\renewcommand{\d}{\text{d}}

\renewcommand{\vec}[1]{{\bf{#1}}}

\newcommand{\mg}{\texttt{MadGraph5\_aMC@NLO}}

\begin{document}

\providecommand{\abs}[1]{\lvert#1\rvert}

\newcommand{\Znunujets}{(Z\to{\nu\bar{\nu}})+\text{jets}}
\newcommand{\Welnujets}{(W\to{\ell\nu})+\text{jets}}
\newcommand{\Znunujet}{(Z\to{\nu\bar{\nu}})+\text{jet}}
\newcommand{\Welnujet}{(W\to{\ell\nu})+\text{jet}}

%%%%%%%%%%%%%%%%%%%%
\title{Non-linear top-Higgs CP violation}
%%%%%%%%%%%%%%%%%%%%
\begin{abstract}
Searches for additional sources of CP violation at the Large Hadron Collider are a central part
of the Higgs physics programme beyond the Standard Model. Studies employing so-called signed
observables that track CP violation through purpose-built asymmetries bolster efforts based on
Higgs boson rate analyses under clear assumptions. A possibility, which is so far unexplored at the LHC,
is a significant non-linear realisation of CP-violation, which is naturally described in non-linear Higgs Effective
Field Theory (HEFT). We perform an analysis of the HL-LHC potential to constrain such interactions considering
a large range of single and double Higgs production processes, including differential information where this
is statistically and theoretically possible. A particular emphasis of our work is distinguishing expected correlations
in the Standard Model Effective Field Theory from those attainable in HEFT.
\end{abstract}

%%%%%%%%%%%%%%%%%%%%%%%%%%%%%%%%%%
%%%%
\author{Akanksha Bhardwaj}\email{akanksha.bhardwaj@glasgow.ac.uk} 
\affiliation{School of Physics \& Astronomy, University of Glasgow, Glasgow G12 8QQ, UK\\[0.1cm]}
\affiliation{Department of Physics, Oklahoma State University, Stillwater, OK, 74078, USA\\[0.1cm]}
%%%%
\author{Christoph Englert} \email{christoph.englert@glasgow.ac.uk}
\affiliation{School of Physics \& Astronomy, University of Glasgow, Glasgow G12 8QQ, UK\\[0.1cm]}
%%%%
\author{Dorival Gon\c{c}alves}\email{dorival@okstate.edu} 
\affiliation{Department of Physics, Oklahoma State University, Stillwater, OK, 74078, USA\\[0.1cm]}
%%%%
\author{Alberto Navarro}\email{alberto.navarro\_serratos@okstate.edu}
\affiliation{Department of Physics, Oklahoma State University, Stillwater, OK, 74078, USA\\[0.1cm]}
%%%%
%%%%%%%%%%%%%%%%%%%%%%%%%%%%%%%%%%

%%%%%%%%%%%%%%%%%%%%%%%
%%%%%%%%%%%%%%%%%%%%%%%
\maketitle

%%%%%%%%%%%%%%%%%%%%%%%%%%%%%%%%%%
\section{Introduction}
\label{sec:intro}
The interactions of the Higgs boson are generally considered as harbingers
of new interactions beyond the Standard Model (BSM). The precision study
of the Higgs boson at the Large Hadron Collider (LHC) has therefore opened
a new territory in our understanding of the electroweak scale. While the
precise nature of the latter is still unclear, it is reasonable to expect
that whatever the mechanism responsible for electroweak symmetry breaking,
it might have wider ramifications for the as yet unresolved questions of the SM.

In BSM scenarios, such as multi-Higgs extensions, the Higgs boson interactions
can introduce additional sources of CP violation which can address one of the Sakharov
criteria that the SM falls short of~\cite{Basler:2021kgq,Anisha:2022hgv,Goncalves:2023svb}. 
From a theoretical
standpoint, certain Higgs couplings  are more susceptible to  pronounced new physics effects.
For instance, the extensively studied CP-odd Higgs-vector boson interactions can appear only through operators  of dimension-six 
or higher~\cite{Buchmuller:1985jz,Grzadkowski:2010es}, being naturally suppressed by the new physics scale. 
In contrast,  CP-odd Higgs-fermion couplings can already appear at tree level
leading to naturally larger CP violation effects. The top quark Yukawa coupling, owing to its  magnitude, 
plays a crucial role in this discussion and emerges as a particularly sensitive probe for physics beyond the 
SM.

Model-agnostic approaches employing effective 
field theory highlight a range of effective interactions in a coarse-grained dimension-six approach
 that have been scrutinized in a range of experimental analyses at the LHC so far.
In particular, additional (C)P violation in the top-Higgs sector 
\begin{equation}
\label{eq:cpphase}
\sim i \,\bar t \gamma^5 t \,h
\end{equation} 
can be constrained in gluon fusion~\cite{Plehn:2001nj,Hankele:2006ma,Klamke:2007cu} and top-Higgs production~\cite{Buckley:2015vsa,Boudjema:2015nda,Demartin:2014fia,Casolino:2015cza}.\footnote{Approaches to disentangle these top-Yukawa interaction modifications
from $\sim G_{\mu\nu} \tilde G^{\mu\nu} h$ contact interactions have been discussed in~\cite{Englert:2019xhk}.} The relevance of CP-violating Yukawa interactions for 
low-energy precision dipole measurements have been revisited recently in Ref.~\cite{Brod:2023wsh}.

One way of pinning down such interactions phenomenologically at hadron colliders 
is the construction of asymmetric observables, which then serve as strong tests of CP-violation without 
relying on CP-even rate information such as cross sections or transverse momentum spectra. However, for some processes,
the expected rate even at 3/ab of the high-luminosity (HL) LHC phase is too limited to construct statistically sensible asymmetries. 
In addition, some processes, e.g. involving scalar final states, do not show interference-related asymmetries. Either case then warrants their
inclusion under the hypothesis that no additional sources of new physics are present, relying on simple hypothesis testing.

In this work, we ask the question of how sensitive the LHC can
be to sources of \emph{non-linear CP violation}. While Ref.~\cite{Englert:2019xhk} discusses
approaches to disentangle gluonic from top-philic sources, the question of how correlated 
CP violation across different Higgs multiplicities remains open. Such freedom becomes apparent within the context of Higgs Effective Field Theory (HEFT) when contrasted with correlations expected within the  Standard Model Effective Field Theory (SMEFT)~\cite{Grzadkowski:2010es,Dedes:2017zog}. This possibility also opens up a novel avenue to decouple dipole moment constraints from TeV scale investigations. As shown in~\cite{Brod:2023wsh}, dipole constraints are highly constraining when considering exclusively the interaction of Eq.~\eqref{eq:cpphase}, but can be significantly relaxed when considering analogous CP violation for light flavour Higgs interactions. This comes at the price of a loss of phenomenological sensitivity, as such Higgs interactions are phenomenologically not always accessible at the LHC. CP violation measured in low energy dipole measurements dominantly sourced in di-Higgs interactions would be further loop and light-flavour Yukawa suppressed and will therefore be less constrained.

This note is organized as follows: In Sec.~\ref{sec:heftcp}, we introduce the interactions studied in this work. Particular emphasis is given 
to the distinctive patterns of CP violation predicted in SMEFT as opposed to the more general HEFT parametrisation. The accurate discrimination of non-linear CP violation requires a robust statistical handle on single Higgs production processes, serving as a prerequisite for the subsequent utilization of di-Higgs production to effectively constrain non-linearity. The processes and the assumptions under which they are included in this work are detailed in Sec.~\ref{sec:procs}. Sec.~\ref{sec:nonlincp} is devoted to the discussion of our fit to non-linear CP violation. We summarize in Sec.~\ref{sec:conc}.

%%%%%%%%%%%%%%%%%%%%%%%%%%%%%%%%%%
\section{HEFTy CP violation}
\label{sec:heftcp}
%%%%%%%%%%%%%%%%%%%%%%%%%%%%%%%%%%
As alluded to above, within the  SMEFT approach, we consider the
operator
\begin{equation}
\label{eq:op}
{\cal{O}}_{t\Phi} = |\Phi|^2 \bar Q_L \Phi^c t_R \,.
\end{equation}
$\Phi$ denotes the Higgs doublet, $\Phi^c = i\sigma^2 \Phi^\ast$, and $Q_L, t_R$ are the left and right-chiral fermion doublet and singlet relevant for the top interactions, respectively.
This operator leads in the broken phase to P-violating interactions for
complex Wilson coefficients.
Therefore, signs of CP-violation across different Higgs multiplicities are correlated as a consequence of
the $SU(2)$ doublet structure of the Higgs boson. For instance, the CP-violating tree-level three and four-point irreducible vertex functions obey
\begin{equation}
{\Gamma_{\bar t t h}\over \Gamma_{\bar t t h^2} }\bigg|_{\gamma^5,\text{SMEFT}}= {v \over 3}\,, 
\end{equation}
with $v\simeq 246~\text{GeV}$ as the vacuum expectation value of the Higgs field.
In this context, although additional sensitivity from CP-sensitive observables in $t \bar{t} hh$ production are welcome, CP-violation
in the top Higgs sector under the assumptions of SMEFT should manifest themselves predominantly in single Higgs physics, which provides the most significant statistical pull in a global analysis.

A phenomenologically identical parametrisation of Eq.~\eqref{eq:op}, which we will use in the following, is given by
\begin{equation}
{\cal{L}}^{\text{SMEFT}}_{\alpha,1} = -\frac{m_t}{v} \, \kappa_{t} \,\bar{t}(\cos\alpha + i \gamma^5 \sin \alpha)\,t \,h \,.
\label{eq:smeft}
\end{equation}
Here, $\alpha$ represents the CP-phase and $ \kappa_{t}$ is a real number that determines the strength of the interaction. In this parametrization, the SM is characterized by $\kappa_{t} =1$ and $\alpha=0$. Conversely, for a purely CP-odd interaction, $\alpha$ would be equal to $\pi/2$. This parametrization can be identified with Eq.~\eqref{eq:op} (after renormalisation of the SM Yukawa couplings and assuming a purely CP-even SM coupling of the top quark)
\begin{equation}
{1\over \Lambda^2}
\left(\begin{matrix}
 {\text{Re}\,{C_{t\Phi}}} \\ 
 {\text{Im}\,{C_{t\Phi}}} 
 \end{matrix} \right)
 = - {\sqrt{2}\, m_t\over v^3}  
\left(\begin{matrix}
 \kappa_t  \cos\alpha - 1 \\ 
\kappa_t   \sin\alpha
 \end{matrix}
 \right)\,.
\end{equation}
This directly leads to
\begin{equation}
{\cal{L}}_{\alpha,2}^{\text{SMEFT}} \supset - {3m_t\over 2v^2} \, \bar t ( \{\kappa_t  \cos\alpha-1\}  + i \kappa_t  \gamma^5 \sin\alpha )\, t \, h^2\,,
\end{equation}
which also shows that the $ t\bar t h h$ interactions vanish for the SM point, $(\kappa_t,\alpha)_\text{SM}=(1,0)$. 

Turning to HEFT, which highlights the Higgs boson as a custodial singlet~\cite{Appelquist:1980vg,Longhitano:1980iz,Longhitano:1980tm,Feruglio:1992wf,Appelquist:1993ka,Brivio:2013pma,Brivio:2016fzo}, the top quark mass arises from the non-linear sigma model of
$SU(2)_L\times SU(2)_R\to SU(2)_V$ that can be parametrized as
\begin{equation}
	U(\pi^{a}) = \exp\left({i \pi^{a} \tau^{a}/v}\right)\,,
\end{equation}
with $SU(2)$ generators $\tau^a$ and Goldstone fields $\pi^a$. This field transforms under general $SU(2)_L\times SU(2)_R$ transformations 
as $U\to L\,UR^\dagger$ so that the top quark mass arises from 
\begin{equation}
\label{eq:fermflare}
	{\cal{O}}_{\bar t t }= - %{y_t\, v\over \sqrt{2}} 
	m_t\,\bar Q_L U t_R  \,.
\end{equation}
Owing to the singlet character of the Higgs boson in HEFT, this operator can be dressed with a ``flare'' function
\begin{equation}
Y_t(h)= 1+ c^{(1)} {h\over v} + c^{(2)} {h^2\over 2v^2} + \dots \,,
\end{equation}
suppressing higher monomials of the singlet Higgs field, which are phenomenologically not relevant. 
This leads to CP-violating effects
analogous to ${\cal{L}}_{\alpha}$
\begin{multline}
{\cal{L}}_\text{HEFT} \supset -\frac{m_t}{v} \, \kappa_{t} \,\bar{t}(\cos\alpha + i \gamma^5 \sin \alpha)\,t \,h \\
-{m_t\over 2 v^2}  \, \kappa_{tt}  \, \bar t(\cos\beta + i \gamma^5 \sin\beta)\, t \, h^2\,.
\label{eq:heft}
\end{multline}
However, it is important to note a significant exception: the Higgs multiplicities remain uncorrelated in this context. 
The expressions for $c^{(1)}$ and $c^{(2)}$ become
\begin{equation}
 c^{(1)}= \kappa_t \,e^{i\alpha}\,,~ c^{(2)}= \kappa_{tt} \,e^{i\beta}\,.
\end{equation}
The relative strength of CP-violation for the three and four-point interactions is now characterized by
\begin{equation}
{\Gamma_{\bar t t h}\over \Gamma_{\bar t t h^2} }\bigg|_{\gamma^5,\text{HEFT}} =
{\kappa_t \over \kappa_{tt}}\,{\sin\alpha \over \sin \beta} \, v
\,,
\end{equation}
where the SMEFT trajectory can be recovered by the HEFT choices
\begin{equation}
\begin{split}
\kappa_{tt}^2 &= 9 (1-2\kappa_t\cos\alpha +\kappa_t^2)\,,\\
\tan\beta &= {\kappa_t \sin\alpha \over \kappa_t\cos\alpha -1}\,.
\end{split}
\label{eq:identify}
\end{equation}

CP measurements at ATLAS and CMS are typically carried by constructing asymmetries or ``signed'' observables which isolate interference effects between new physics
and SM contributions. Writing the amplitude of the scattering process ${\cal{M}}={\cal{M}}_\text{SM} + {\cal{M}}_{\cal{O}}$, with $ {\cal{M}}_{\cal{O}}$ denoting the BSM part, the partonic cross sections scale
as 
\begin{equation}
{\text{d}\sigma \over  \text{dLIPS}}\sim  | {\cal{M}}_\text{SM} | + 2\text{Re}({\cal{M}}_\text{SM} {\cal{M}}_{\cal{O}}^\ast) +  | {\cal{M}}_{\cal{O}} |^2\,.
\label{eq:expand}
\end{equation}
Squared CP-odd contributions manifest in CP-even distributions, such as cross sections, transverse momentum distributions, etc. The interference effects
between SM and new physics cancel in these CP-even distributions and are resolved through purpose-built observables. However, for processes with limited statistics, achieving a binned distribution might not always be attainable, even during the high-luminosity phase of the Large Hadron Collider (HL-LHC). We detail the processes we include in our study in the next Sec.~\ref{sec:procs}.

%%%%%%%%%%%%%%%%%%%%%%%%%%%%%%%%%%
\section{Sensitive Processes and details of the analysis}
%%%%%%%%%%%%%%%%%%%%%%%%%%%%%%%%%%
\label{sec:procs}
%----------

%----------
\subsection{Inclusive $gg\to h$ production}
%----------
Gluon fusion Higgs production has become one of the standard candles to study electroweak symmetry breaking at the LHC ever since the Higgs boson's discovery. The phenomenological precision programme is well underway and the experiments have laid out a detailed roadmap towards the HL-LHC phase. When rate information is considered, the cross section and decay widths are known to provide important handles on potential CP violation (see, e.g., Refs.~\cite{Djouadi:2005gi,Djouadi:2005gj}). To reflect the sensitivity of this process to phases of Yukawa interactions as discussed above, we employ  the ECFA extrapolation by CMS outlined in Ref.~\cite{CMS:2017cwx}. Specifically, we consider the $h \to \gamma \gamma$ and $h\to ZZ$ signal strength extrapolations, which forecast a sensitivity at 95\% CL of
\begin{align}
{\delta \mu \over \mu}(gg\to h \to \gamma\gamma) &= 3.3\%\,, \\
{\delta \mu \over \mu}(gg\to h \to ZZ) &= 4.6\%\,.
\end{align}
We also include $h\to \tau \tau$ based on an extrapolation of Ref.~\cite{CMS:2017zyp} which sets
\begin{align}
{\delta \mu \over \mu}(gg\to h \to \tau\tau) &= 9.7\%\,.
\end{align}
This aligns with the ECFA projection presented in Ref.~\cite{CMS:2017cwx}. To achieve this, we use \mg~to interpolate the cross section, using a model generated with {\tt{FeynRules}}~\cite{Alloul:2013bka}, {\tt{NLOCT}}~\cite{Degrande:2014vpa}, and {\tt{UFO}}~\cite{Degrande:2011ua} in the finite top mass limit. This interpolation accounts for various coupling choices and is then reweighed based on the SM result to reflect higher-order QCD corrections~\cite{Djouadi:1991tka,Anastasiou:2015vya,LHCHiggsCrossSectionWorkingGroup:2011wcg}. Throughout this work, we take into account the modifications of the Higgs branching ratios due to the modified top-Yukawa couplings.

%----------
\subsection{Gluon fusion $h+2j$ production}
%----------
 %---------------------
\begin{figure}[!t]
   \centering
   \includegraphics[width=0.45\textwidth]{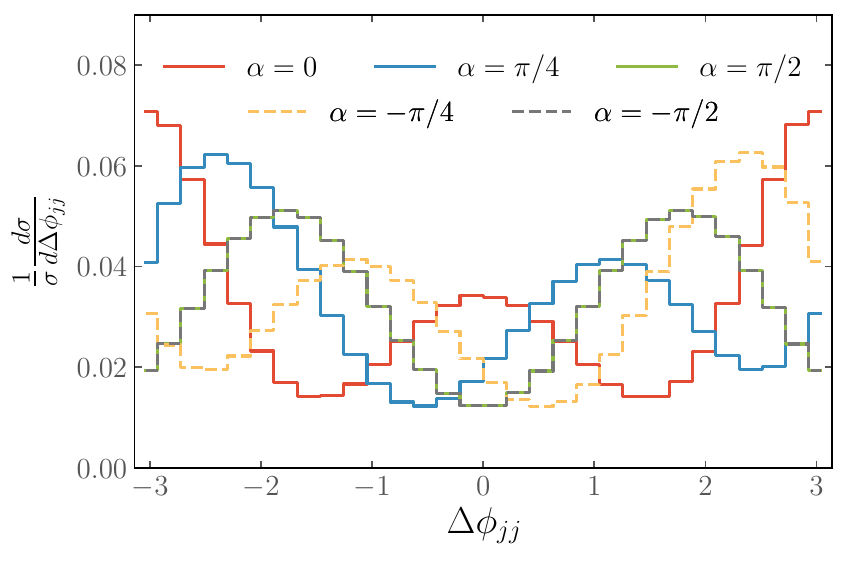}
    \caption{Distribution for the azimuthal angle difference between the two tagging jets $\Delta\phi_{jj}$, as defined in Eq.~\eqref{eq:phijj},  specifically for the $h+ 2j$ sample.}
    \label{fig:hjj}
\end{figure}
%---------------------
The production of a single Higgs boson in association with two jets is a sensitive process due to the introduction of the `signed' angular separation between the tagging jets~\cite{Plehn:2001nj,Klamke:2007cu,Buschmann:2014twa}. Ordering the jets in rapidity $\eta_{j1}>\eta_{j2}$, the azimuthal angular difference 
\begin{equation}
\Delta\phi_{jj} = \phi_{j1}-\phi_{j2}\,
\label{eq:phijj}
\end{equation}
leads to a characteristic angular modulation, which can be exploited to set constraints on
the involved CP-odd interactions. This renders $h+2j$ as a prime candidate for constraining  the single Higgs property 
as compared to the non-linear deviations.\footnote{Gluon fusion of Higgs pairs in association with two jets has been studied in
Ref.~\cite{Dolan:2013rja} and faces significant phenomenological challenges at the LHC. Therefore, we will not discuss this process further.} Therefore, this process
has been used relatively early in the LHC Higgs programme to set constraints on sources of CP violation.  

For our analysis, we use these ATLAS results as a baseline for extrapolation~\cite{ATLAS:2018hxb}. We employ the {\tt{Vbfnlo}}~\cite{Arnold:2008rz,Baglio:2011juf} Monte Carlo to 
include the finite top-mass effects that shape the phenomenology of the $h+2j$ final state, including the phase of the top Yukawa interaction. For illustrational purposes, we present the SM and new physics $\phi_{jj}$ distributions in Fig.~\ref{fig:hjj}. We extract efficiencies for a SM sample mapped onto the results of~\cite{ATLAS:2018hxb} and generalize these to the BSM parameter choices involving CP-odd contributions, following the procedure detailed in \cite{Englert:2019xhk}.

%----------
\subsection{Top-associated Higgs production $t\bar{t}h$}
%----------

%---------------------
\begin{figure*}[!t]
   \centering
   \includegraphics[width=0.45\textwidth]{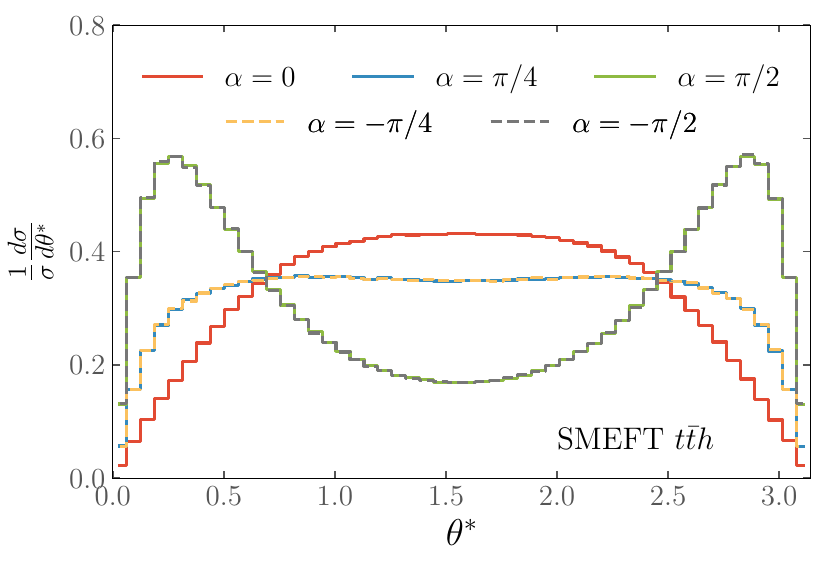}
   \includegraphics[width=0.45\textwidth]{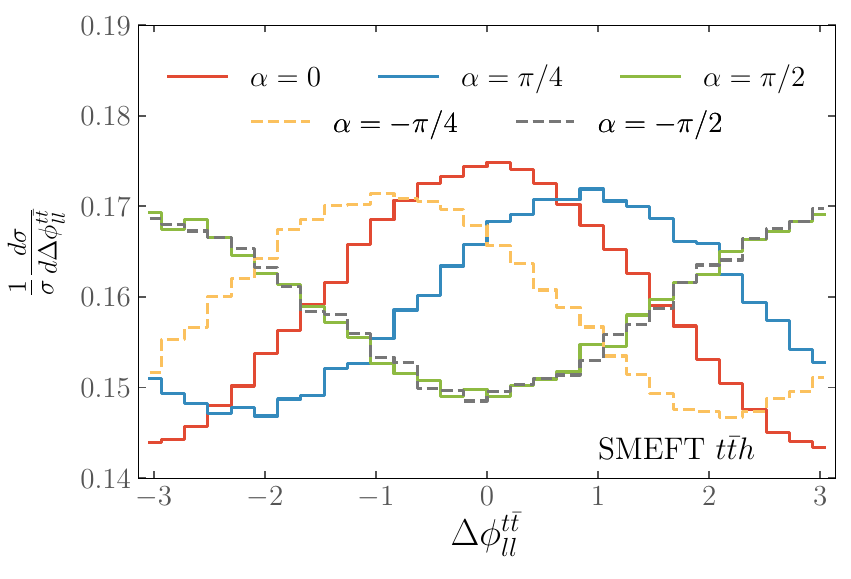}
    \includegraphics[width=0.45\textwidth]{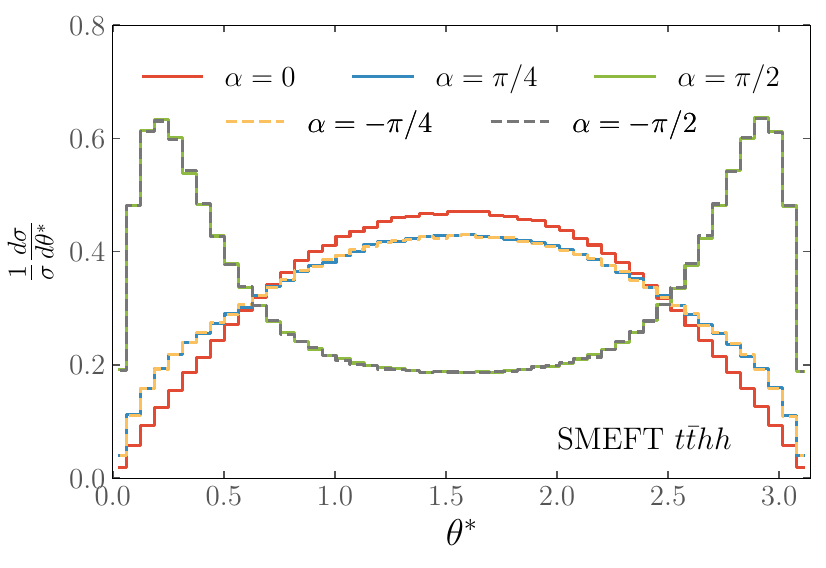}
   \includegraphics[width=0.45\textwidth]{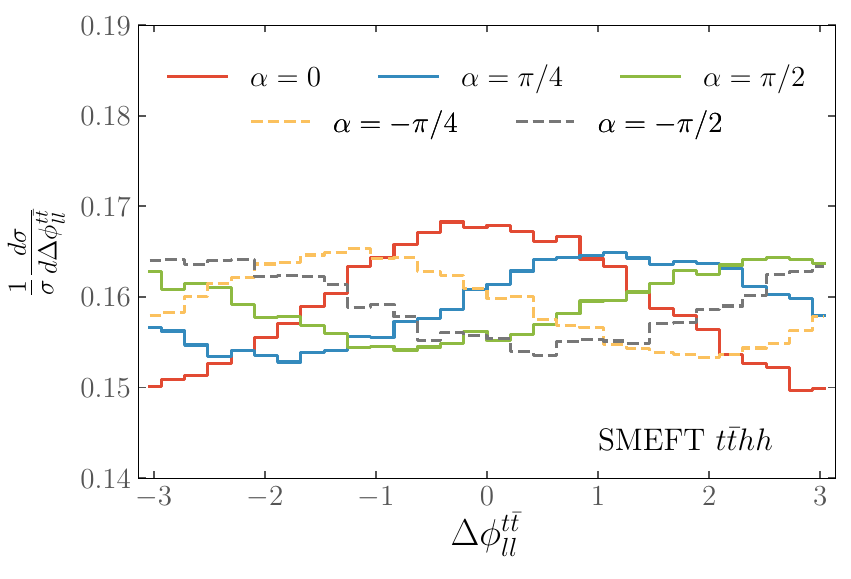}
    \caption{Collins-Soper angle $\theta^*$ (left) and azimuthal angle distribution $\Delta \phi_{\ell\ell}^{t\bar{t}}$
     (right) for $t\bar{t}h$  (top) and  $t\bar{t}hh$ (bottom) processes with dileptonic top pair final state. We consider the SMEFT framework for demonstration purposes.}
    \label{fig:tth}
\end{figure*}
%---------------------

The $pp\to t\bar{t}h$ channel plays a crucial role in probing the Higgs-top CP-structure at the tree-level,
disentangling possible new physics effects~\cite{Ellis:2013yxa,Boudjema:2015nda,Buckley:2015vsa,Buckley:2015ctj,Gritsan:2016hjl,Goncalves:2016qhh,Mileo:2016mxg,AmorDosSantos:2017ayi,Azevedo:2017qiz,Goncalves:2018agy,ATLAS:2018mme,CMS:2018uxb,Bortolato:2020zcg,Cao:2020hhb,Martini:2021uey,Goncalves:2021dcu,Bahl:2021dnc,Azevedo:2022jnd,Ackerschott:2023nax}. Several  kinematic observables have been proposed in
the literature to investigate the CP structure of the Higgs-top interaction in this channel. Among
those, the Collins-Soper angle $\theta^*$, which is the angle between the top quark and the beam direction in the $t\bar{t}$
CM frame, features as one of the most sensitive observables to CP at the non-linear level~\cite{Goncalves:2018agy,Goncalves:2021dcu} (in the sense of Eq.~\eqref{eq:expand}). Genuine CP-odd observables
can also be defined exploiting the top-quark polarization that is carried over to its decay products. 
It is possible to form
tensor products involving the top quark pair and their decay products, represented as 
$\epsilon (p_t, p_{t}, p_i, p_k) \equiv  \epsilon_{\mu\nu\rho\sigma} p_t^\mu p_{\bar t}^\nu p_i^\rho p_k^\sigma$~\cite{Goncalves:2018agy,Barman:2021yfh}. 
This tensor product can be simplified as $\vec p_t \cdot (\vec p_i\times \vec p_k)$ in the $t\bar{t}$ CM frame and provides 
a basis for defining azimuthal angle differences that exhibit an odd behavior under CP transformations
\begin{align}
        \!\!\! \Delta \phi_{ik}^{t\bar{t}} \!=\!
    \text{sgn} \left[\vec{p}_{t} \!\cdot\! (\vec{p}_{i} \!\times\! \vec{p}_{k})\right] 
   \arccos \!\left( \frac{\vec{p}_{t} \!\times\! \vec{p}_{i}}{|\vec{p}_{t} \!\times\! \vec{p}_{i}|} \!\cdot\! \frac{\vec{p}_{t} \!\times\! \vec{p}_{k}}{|\vec{p}_{t} \!\times\! \vec{p}_{k}|}\right)\!.
    \label{eq:tth_CP_odd}
\end{align}
We present  both the Collins-Soper $\theta^*$  and the azimuthal angle distribution $\Delta \phi_{\ell\ell}^{t\bar{t}}$ for  dileptonic top pair final states in the top panel of  Fig.~\ref{fig:tth}. The $t\bar{t}hh$ channel, which we will discuss further below, may provide another complementary avenue to probe the Higgs-top coupling at the tree-level~\cite{Englert:2014uqa}. Observables that mirror those defined for the $t\bar{t}h$ process can also be established for this additional channel as presented in the bottom panel of Fig.~\ref{fig:tth} (see also~\cite{Liu:2015aka}).

We extract the direct Higgs-top CP sensitivity at the HL-LHC from our previous analysis in Ref.~\cite{Barman:2021yfh}. In this study, we employ a synergy of machine learning techniques and streamlined kinematic reconstruction methods to enhance the new physics sensitivity, exploring the complex $t\bar{t}h$ multi-particle phase space. The analysis encompasses a range of final states, including hadronic, semi-leptonic, and di-leptonic top pair decays, all in conjunction with the Higgs decay $h\to \gamma\gamma$. It is noteworthy that the experimental projections from ATLAS and CMS  indicate that the $h\to \gamma\gamma$ final state will display the dominant sensitivities to the $t\bar{t}h$ channel  at the HL-LHC~\cite{Cepeda:2019klc}.

%----------
\subsection{$Z$ boson-associated Higgs production}
%----------
Although the leading contribution for the Higgstrahlung channel $Zh$ arises at tree level with
 $q\bar{q}\to Zh$, this channel displays relevant $\mathcal{O}(\alpha_s^2)$ corrections through 
 the loop-induced gluon fusion $gg\to Zh$~\cite{Altenkamp:2012sx,Harlander:2013mla}, which are particularly important in the boosted regime, 
 $p_{Th}\sim m_t$~\cite{Englert:2013vua,Goncalves:2015mfa,Hespel:2015zea,Rossia:2023hen}. Setting limits in these exclusive phase-space regions 
 is an experimental challenge and to obtain a qualitative sensitivity estimate, we perform a more detailed signal
 vs. background investigation. 

We denote the $q\bar q$ and $gg$ subprocesses as $Zh_{\rm DY}$ and $Zh_{\rm GF}$, respectively. 
The $Zh_{\rm GF}$ process exhibits sensitivity to the linear and quadratic terms of the top-Higgs Yukawa
 coupling. Owing to the large destructive interference for the top Yukawa terms, the $Zh_{\rm GF}$ 
 contribution can be sensitive to the magnitude and sign of a possible non-standard top-Higgs coupling 
 $(\kappa_t,\alpha)$.\footnote{A comprehensive study of the angular moments for the $Z$ boson in the 
 $Zh_{\rm GF}$ channel is presented in Appendix~\ref{sec:zh-appendix}. These probes work as additional 
 analyzers for the Higgs-top CP violation effects.}

We now investigate the sensitivity to new physics in the ${gg\to Z(\ell\ell)h(b\bar{b})}$ channel. Our signal 
comprises two charged leptons, $\ell=e$ or $\mu$, reconstructing a boosted $Z$ boson recoiling against 
two $b$-jets. The main background processes are $Zb\bar{b}$, $t\bar{t}$+jets, and $ZZ$.
For our analysis, we generate the signal sample $Zh_{GF}$ using \mg~\cite{Alwall:2014hca, Hirschi:2015iia}, 
 while the background samples are simulated with Sherpa+OpenLoops~\cite{Gleisberg:2008ta,Cascioli:2011va,Denner:2016kdg}, 
 following the study presented in Ref.~\cite{Goncalves:2018fvn}. The $Zh_{\mathrm{DY}}$, $Zb\bar{b}$, and $ZZ$ 
 background samples are merged at LO with up to one additional jet emission using the CKKW 
 algorithm~\cite{Catani:2001cc, Hoeche:2009rj}. We normalize their cross sections to the NLO rates 
 obtained from Ref.~\cite{Goncalves:2015mfa}. Additionally, we generate the $t\bar{t}$ background at NLO 
 using the MC@NLO algorithm~\cite{Frixione:2002ik,Hoeche:2011fd}, considering hadronization and underlying 
 event effects in our simulation.

To reconstruct the signal events, we require two same-flavor leptons with opposite-sign charges satisfying 
${p_{T\ell}>30}$~GeV and $|\eta_\ell |<2.5$, within the invariant mass range ${75~\text{GeV}<m_{\ell\ell}<105}$~GeV. 
The $Z$ boson is required to have a large boost, $p_{\mathrm{T}\ell\ell}>200$GeV. We adopt the BDRS analysis for 
the $h\to b\bar{b}$ tagging~\cite{Butterworth:2008iy}, which involves re-clustering the hadronic activity using 
the Cambridge-Aachen jet algorithm~\cite{Cacciari:2011ma} with ${R=1.2}$. We impose at least one boosted 
fat-jet with $p_{TJ}>200$~GeV and $|\eta_{J}|<2.5$, Higgs-tagged using the BDRS algorithm, which demands 
three sub-jets with the two leading sub-jets being $b$-tagged. We assume a flat 70\% $b$-tagging efficiency and 
a 1\% mis-tag rate. To further improve the signal-to-background ratio, we impose a constraint on the filtered Higgs 
mass within the range ${|m_{h}^\text{BDRS}-m_h|<10}$~GeV, where $m_h^{}=125$~GeV. The resulting event rate
 is presented in Tab.~\ref{tab:analysis_zh}.

%-------------------------------------------------------
\begin{table}[h!]
\centering
\begin{tabular}{l  | c | c | c | c | c  }
 \multirow{1}{*}{} &
 \multicolumn{1}{c|}{$Zh_{\mathrm{GF}}$}  &
 \multicolumn{1}{c|}{$Zh_{\mathrm{DY}}$} &
  \multicolumn{1}{c|}{$Zb\bar{b}$} &
 \multicolumn{1}{c|}{$t\bar{t}$}  &
  \multicolumn{1}{c}{$ZZ$}   \\
  \hline
{BDRS reconstruction} & \multirow{2}{*}{0.03} & \multirow{2}{*}{0.16} & \multirow{2}{*}{0.35} & \multirow{2}{*}{0.02} & \multirow{2}{*}{0.02} \\
$|m_{h}^{\rm BDRS}-m_h|<10$~GeV  &  &  &   &   &  \\
\end{tabular}
  \caption{Signal rate for $Zh_{\mathrm{GF}}$ and  background rates for $Zh_{\mathrm{DY}}$, $Zb\bar{b}$, $t\bar{t}$, and $ZZ$. The signal is generated for the SM hypothesis $(\kappa_t,\alpha)_\text{SM}=(1,0)$ and rates are given in fb after the BDRS analysis.}
\label{tab:analysis_zh}
\end{table}
%-------------------------------------------------------

%----------
\subsection{Beyond linearity: $t\bar thh$ inclusive $hh$ production}
%----------
We now turn to the discussion of processes that provide genuine sensitivity to non-linearity via the production of final states containing a pair of Higgs bosons. Such processes are statistically limited
at the LHC, yet in the case of gluon fusion production $gg\to hh$ relatively well understood, both theoretically and experimentally. In particular, Higgs pair production has
been subject to considerable experimental scrutiny already, and  detailed experimental forecasts for the HL-LHC frontier have
been made available, similar to the case of $gg\to h$ production. To this end, we consider the $b\bar b \gamma\gamma + b\bar b \tau \tau$ extrapolation of \cite{ATLAS:2022faz}  
\begin{equation}
\label{eq:gghhx}
{\sigma(hh) \over \sigma(hh)_{\text{SM}}}< 2 \,,
\end{equation}
at 95\% confidence level (CL), which could lower to $1.1$ if systematics become sufficiently well-controlled. Both $b\bar b \tau \tau$ and $b\bar b \gamma \gamma$ have comparable statistical sensitivity
and we include them on an equal footing to our statistical analysis, again taking into account the effect of modified Higgs branching ratios as a function of $(\kappa_t,\alpha)$.
Similar to the $gg\to h$ process, we interpolate $gg\to hh$ production using \mg~in the finite top mass limit to reflect the constraint from Eq.~\eqref{eq:gghhx} within our combined analysis in Sec.~\ref{sec:nonlincp}.

%---------------------
\begin{figure}[!b]
   \centering
   \includegraphics[width=0.45\textwidth]{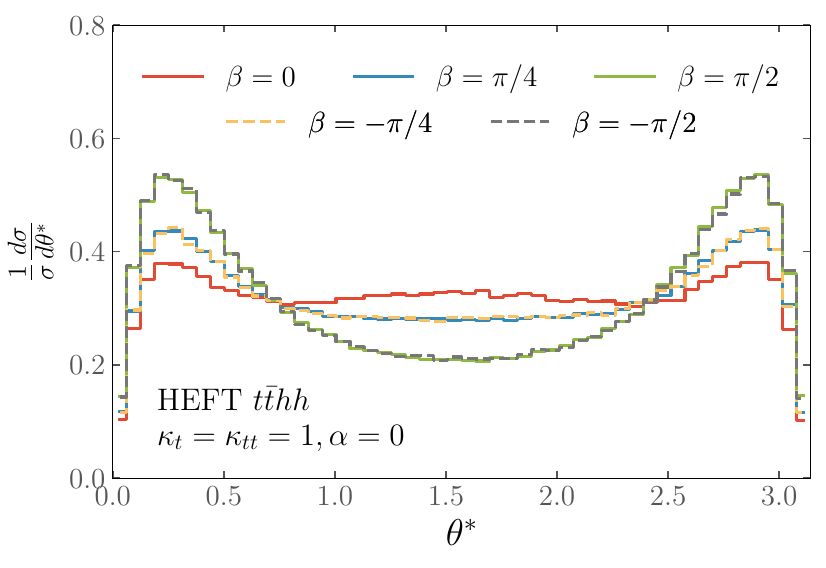}
    \caption{Collins-Soper angle $\theta^*$ for the $t\bar{t}hh$ process in the HEFT framework. We use $\kappa_t=\kappa_{tt}=1$ and $\alpha=0$ for illustration.}
    \label{fig:tthh}
\end{figure}
%---------------------

In comparison, the $ t\bar t hh$ process is rather more complex and currently only proof-of-principle analyses exist, e.g., Refs.~\cite{Englert:2014uqa,ATLAS:2016ggl} for the HL-LHC. The former predicts around 10 signal events in the SM for a $b$-rich final state. Being statistically limited, shape analyses of signed observables, which can be constructed similar to the $ t\bar t h$ process, will not yield relevant exclusion constraints. Selected, relevant observables for this channel  are illustrated in Fig.~\ref{fig:tth} (bottom panel) and Fig.~\ref{fig:tthh} within the SMEFT and HEFT frameworks, respectively. Given the statistical limitation, we incorporate the 95\%~CL cross section exclusion limit for $ t\bar t hh$ 
\begin{equation}
{\sigma(t\bar t hh) \over \sigma( t\bar t hh)_{\text{SM}}}< 1.4 ... 6.8
\end{equation}
based on the analyses of Refs.~\cite{Englert:2014uqa,ATLAS:2016ggl}. This limit does not include the impact of background systematics, which can reduce this estimate. However, it is worth highlighting that within the experimental context, the potential to improve this channel remains relatively unexplored. 
We note that the cross section is driven by the four-point interactions~\cite{Banerjee:2019jys}, similar to $gg\to hh$,  and has been a focus of studies like for the composite Higgs framework~\cite{Grober:2010yv}.

%%%%%%%%%%%%%%%%%%%%%%%%%%%%%%%%%%
\section{A fit to non-linear CP violation in the top-Higgs sector}
\label{sec:nonlincp}
%%%%%%%%%%%%%%%%%%%%%%%%%%%%%%%%%%
The asymmetries and total rates are used to set CL limits on the parameter space $(\kappa_t, \alpha,\kappa_{tt},\beta)$, assuming the SM as the null hypothesis. To this end, we consider a $\chi^2$ statistic defined as
\begin{equation}
	\label{eq:chi2}
	\chi^2 = \sum_i {(N_i - N_i^\text{SM})^2\over \sigma^2_{i}}\,.
\end{equation}
Here, the index $i$ runs over a binned distribution where this is statistically warranted, or $i=1$ for constraints from cross sections. $N_i$ denotes
the event count in a particular bin (or the entire signal event count for cross sections) for a given luminosity, which we set to ${\cal{L}}=3~\text{ab}^{-1}$. We tune the uncertainties $\sigma^2_{i}$ to reproduce the quoted single channel sensitivities. Given these individual $\chi^2$ contributions, we can then consider their combination, increasing the degrees of freedom depending on the hypothesis under investigation.

%---------------------
\begin{figure}[!t]
   \centering
   \includegraphics[width=0.45\textwidth]{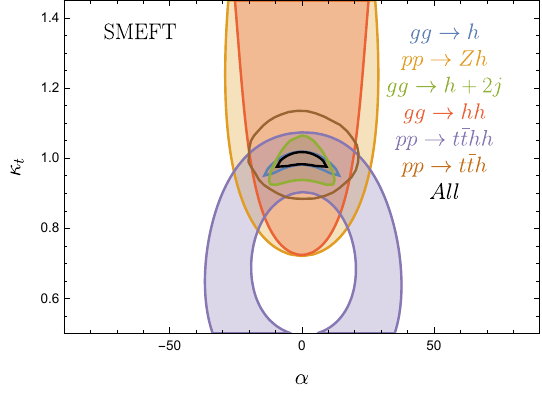}
    \caption{$95\%$ CL limits on $(\kappa_t,\alpha)$ for the 13 TeV HL-LHC with $3~\text{ab}^{-1}$ in the SMEFT framework, for which $(\kappa_{tt},\beta)$ are related to the former via Eq.~\eqref{eq:identify}.}
    \label{fig:tth_all}
\end{figure}
%---------------------

Before we turn to combinations and the comparison between SMEFT and HEFT, it is instructive to highlight the sensitivity of each of these channels and how multi-Higgs production serves as means to distinguish non-linearity. We will focus on the HL-LHC data set in the following. 

In Fig.~\ref{fig:tth_all}, we show the sensitivity of all channels before combination, focusing on the SMEFT parametrization that singles out the correlation of Eq.~\eqref{eq:identify} 
across the different Higgs multiplicities. As expected, the most sensitive channels in SMEFT are those with highest statistical abundance. Under the assumption of suppressed competing coupling modification in SMEFT, this is given by the inclusive gluon fusion rate along the $\kappa_t$ direction. Exploitable angular correlations in the $h+2j$ mode augment the sensitivity along the direction of the CP angle.\footnote{This information is also accessible at the interference level in Eq.~\eqref{eq:expand} and is therefore  relatively robust with regard to linearizing differential cross sections in the EFT expansion.} Directly probing the top Yukawa coupling through the $t\bar{t}h$ channel also leads to relevant complementary constraints.

%---------------------
\begin{figure}[!t]
   \centering
   \includegraphics[width=0.45\textwidth]{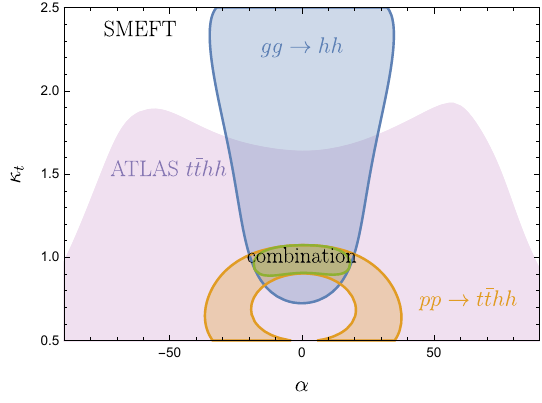}\\[0.3cm]
   \includegraphics[width=0.45\textwidth]{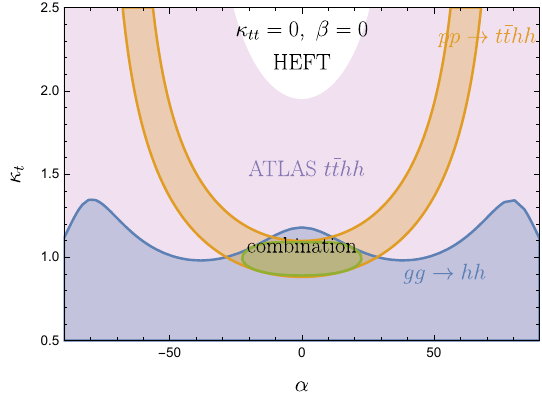}
    \caption{ $95\%$ CL limits on the $(\kappa_t,\alpha)$ plane at the 13 TeV HL-LHC with $3~\text{ab}^{-1}$ of data for $hh$ and  $t\bar{t}hh$ channels in SMEFT (top) and HEFT (bottom) frameworks with $(\kappa_{tt}=0,\beta=0)$. We also highlight the importance of the $t\bar t hh$ channel in collapsing the available parameter space, a non-trivial combination is shown for the stringent $t\bar t hh$ assumption.}
    \label{fig:smeft_tthh}
\end{figure}
%---------------------

Given the reduced sensitivity in the multi-Higgs production, the $(\kappa_t,\alpha)$ constraints carry over from the SMEFT parametrization to HEFT, modulo changes in the number of degrees of freedom and the small pull provided by the SMEFT correlation in light of the correlation of Eq.~\eqref{eq:identify}. The importance of the latter correlation becomes clear when contrasting the $gg\to hh$ and $t\bar t hh$ combination in SMEFT against HEFT for $\kappa_{tt}=\beta=0$, as depicted in  Fig.~\ref{fig:smeft_tthh}. This comparison highlights the relevance of quartic $t\bar thh$ contact interactions for these final states. As can be seen, these particular BSM contact interactions  drive the cross section for the di-Higgs production modes.

\begin{figure}[!t]
   \centering
   \includegraphics[width=0.45\textwidth]{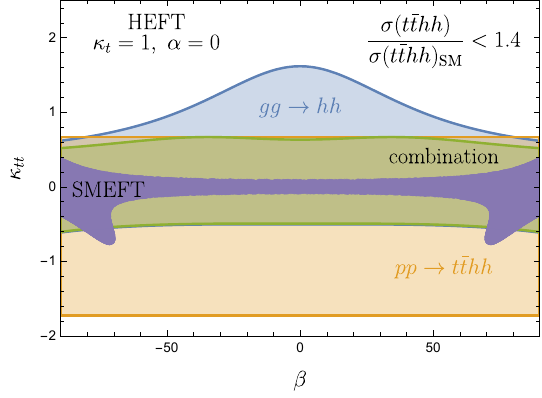}
    \caption{ $95\%$ CL limits on $(\kappa_{tt},\beta)$ at the 13 TeV HL-LHC with $3~\text{ab}^{-1}$ of data for $hh$ and  $t\bar{t}hh$ channels in the HEFT framework with $(\kappa_{t}=1,\alpha=0)$. The SMEFT region selected from a fit to single Higgs data is highlighted for comparison.}
    \label{fig:heft_ktt_beta}
\end{figure}

Assuming a SM value in HEFT for the single Higgs modes, $(\kappa_t,\alpha)=(1,0)$, the expected constraints from purely non-linear CP violation are given in Fig.~\ref{fig:heft_ktt_beta}. As can be seen, the multi-Higgs production modes can be used to set constraints mostly on the magnitude of the contact interaction, whilst the expected sensitivity is not large enough to constrain its phase. This blind direction could potentially be explored through the multi-particle final state kinematics as illustrated in Fig.~\ref{fig:tthh}. However, achieving this may necessitate a higher event rate and might realistically only become feasible at upcoming higher-energy colliders, such as the FCC-hh~\cite{Banerjee:2019jys}.
 
SMEFTy extensions close to the decoupling limit select a  subspace of HEFT. Given the correlation predicted by SMEFT-like extensions of the SM, we can therefore employ these production modes to highlight the expected sensitivity for $\kappa_{tt},\beta$ when comparing SMEFT and HEFT in Fig.~\ref{fig:heft_ktt_beta}. The SMEFT contour highlights the correlation of a combined fit of the most sensitive single Higgs channels in SMEFT projected onto the $(\beta,\kappa_{tt})$ plane given the correlations of Eq.~\eqref{eq:identify}. For illustration purposes, we limit the HEFT parameter space to SM couplings in the single Higgs sector (the corresponding couplings will be relatively well measured at 3/ab and the di-Higgs cross sections are predominantly sensitive to the multi-Higgs couplings). Clearly, a SM-like outcome of the single Higgs measurements renders the available parameter space in the di-Higgs couplings relatively limited in SMEFT. Even if the optimistic $t\bar t hh$ constrain is relaxed to looser constraints, $gg\to hh$ production is still sensitive to significant quartic $t\bar t hh$ vertices and associated CP violation in~HEFT. 

When reducing the size of $\kappa_{tt}$, the sensitivity to $\beta$ is naturally suppressed. Higher sensitivity in the relevant channels is therefore key to further maximise the LHC potential: the $t\bar t hh$ contour in Fig.~\ref{fig:heft_ktt_beta} only slightly bends for larger angles $\beta$. Perhaps an unrealistic improvement over the quoted constraints would extend the $\beta$ sensitivity. The role of $t\bar t hh$ production remains critical, even when only the $\kappa_{tt}$ effects are considered. Feasibility studies
beyond a first exploratory studies, e.g.~\cite{ATLAS:2016ggl}, should continue to maximise the value of LHC data. Of course, the statistical limitations
present for multi-Higgs mode at the LHC are naturally relaxed at a future hadron machine such as FCC-hh, envisioned to operate at 100~TeV with a target luminosity of $30~\text{ab}^{-1}$. A more fine grained approach exploiting angular correlations as demonstrated in Fig.~\ref{fig:tth} (bottom panel) and Fig.~\ref{fig:tthh} will become possible, which will lead to a qualitatively new $t\bar t hh$ exclusion.

 %---------------------
%---------------------

%%%%%%%%%%%%%%%%%%%%%%%%%
\section{Conclusions}
\label{sec:conc}
%%%%%%%%%%%%%%%%%%%%%%%%%
In this work, we have examined the potential of the LHC to constrain CP phases of the top-Yukawa interactions combining the sensitivity of a range of single- and double Higgs production
processes. Single Higgs processes encompass all the relevant correlations in dimension-six SMEFT, and multi-Higgs production does not lead to significant sensitivity gain. 
However, this paradigm shifts when considering non-linear sources of CP violation.
Given the limited rates of multi-Higgs production at the LHC, the resulting constraints are naturally less stringent than those anticipated from single Higgs physics, especially when incorporating rate information under appropriate assumptions. Nonetheless, the LHC shows sensitivity, in particular when discriminating between SMEFTy and HEFTy CP violation in the top-Higgs sector.

%-------------------------------------------------------
\begin{table*} [!ht]
\centering
\begin{tabular}{c | cccccccc}
\hline \hline
  & $A_0$ & $A_1$ & $A_2$ & $A_3$ & $A_4$ & $A_5$ & $A_6$ & $A_7$ \\ \hline
 SM  & 0.027 $\pm 0.003$ & 0.001 $\pm 0.005$ & -0.976 $\pm 0.005$ & -0.003 $\pm 0.004$ & -0.003 $\pm 0.004$ &  0.000 $\pm 0.004$ & 0.001 $\pm 0.005$ &  -0.003 $\pm 0.005$ \\
  $\alpha=\pi/4$  & 0.016 $\pm 0.003$ & -0.006 $\pm 0.005$ & -0.968 $\pm 0.005$ & 0.002 $\pm 0.004$ & 0.003 $\pm 0.004$ & -0.009 $\pm 0.004$ & 0.000 $\pm 0.005$ & 0.011 $\pm 0.005$ \\ 
 $\alpha=-\pi/4$ & 0.017 $\pm 0.003$ & 0.001 $\pm 0.005$ & -0.962 $\pm 0.005$ & -0.002 $\pm 0.004$ & 0.002 $\pm 0.004$ & -0.002 $\pm 0.004$ & 0.003 $\pm 0.005$ & 0.003 $\pm 0.005$\\
 \hline \hline
\end{tabular}
\caption{Angular coefficients $A_i$ for the loop-induced process $g g\rightarrow \ell^{+}\ell^{-}h$ with minimum selections $p_{T,\ell\ell}>200$~GeV, $75~\text{GeV} < m_{\ell\ell}< 105$~GeV. We derive the angular coefficients $A_i$ in the Collins-Soper frame~\cite{Collins:1977iv}. Uncertainties are derived from Monte Carlo statistics.}
\label{tab:Ai}
\end{table*}
%-------------------------------------------------------

Our work re-advertises the relevance of the $ t \bar t hh$ and inclusive $hh$ sensitivity studies. For scenarios that are more closely related to the HEFT parametrisation, the multi-Higgs rates also play central roles in honing sensitivity to non-linear CP violation. Although these processes suffer from limitations at the LHC and their resulting constraints are relatively weak when compared to SMEFT correlations, they provide unique avenues for probing such interactions, in particular because low energy precision experiments (e.g. dipole measurements) will have reduced sensitivity compared to SMEFT.

\acknowledgments
A.B. and C.E. are supported by the STFC under grant ST/T000945/1. C.E. is supported by the Leverhulme Trust under grant RPG-2021-031 and the IPPP Associateship Scheme. DG and AN thank the U.S.~Department of Energy for the financial support, under grant number DE-SC 0016013.

%%%%%%%%%%%%%%%%%%%%%%%%%
\appendix
\section{CP-violation effects to the $Z$ boson angular moments in the $gg\to Zh$ process}
\label{sec:zh-appendix}
%%%%%%%%%%%%%%%%%%%%%%%%%

The angular moments for the $Z$ boson can be used as analyzers for the Higgs-top CP violation effects in the loop-induced $gg\to Zh$ process. 
In general, the differential cross-section for the described process can be represented as
\renewcommand{\d}{\text{d}}
\begin{align}
& \frac{1}{\sigma}\frac{\d\sigma}{\d\cos{\theta}\d\phi} =\nonumber \\
& \frac{3}{16\pi} [ 1+\cos^2{\theta} 
+\frac{ A_{0} }{2} (1-3\cos^2{\theta} )
+ A_{1} \sin{2\theta} \cos{\phi}    \nonumber  \\
& +\frac{ A_{2}}{2} \sin^2{\theta} \cos{2\phi} 
+A_{3} \sin{\theta} \cos{\phi} 
+A_{4} \cos{\theta} \nonumber  \\
&+A_{5} \sin^2{\theta} \sin{2\phi}
+ A_{6} \sin{2\theta} \sin{\phi} 
+ A_{7} \sin{\theta} \sin{\phi}]  \,,
\label{eq:Ai}
\end{align}
where $\theta$ and $\phi$ denote the polar and azimuthal angles of the $\ell^-$ lepton in the $Z$ boson rest frame. The eight coefficients $A_i$,  $i=0,..,7$, 
correspond to the degrees of freedom for the polarization density matrix of a spin-1 particle. Remarkably, the three coefficients $A_{5,6,7}$ are proportional to 
the relative complex phases of the scattering amplitudes~\cite{Goncalves:2018fvn}. Hence, when associated to depleted strong phase contributions from loop contributions,
these coefficients can be sensitive to truly CP-violation effects.

To extract the angular coefficients $A_i$ from our Monte Carlo simulation, we recognize that Eq.~\eqref{eq:Ai} represents a spherical harmonic decomposition for the differential cross-section, utilizing real spherical harmonics $Y_{lm}(\theta,\phi)$ of order $l\le 2$~\cite{MammenAbraham:2022yxp}. Consequently, we can access the angular coefficients by exploring the orthogonality relations of the spherical harmonics. The angular coefficients are projected out using the following relations
 %----
\begin{align}
&A_0=4-\left <10\cos^2\theta \right >,                &A_1& =\left <5\sin 2\theta\cos\phi \right > ,   \nonumber \\
&A_2=\left <10\sin^2\theta\cos 2\phi\right > ,    &A_3&=\left<4\sin\theta\cos\phi  \right>,  \nonumber \\
&A_4=\left <4\cos\theta \right > ,                       &A_5&=\left<5\sin^2\theta\sin 2\phi  \right> , \nonumber \\
&A_6=\left <5\sin 2\theta\sin\phi \right > ,          &A_7&=\left<4\sin\theta\sin\phi \right> ,
\end{align}
%----
and the weighted normalization is defined as 
 %----
\begin{align}
\left <f(\theta,\phi)\right >\equiv \int_{-1}^{1} \d\cos\theta \int_0^{2\pi}\d\phi  \,\frac{f(\theta,\phi) }{\sigma}\frac{\d\sigma}{\d\cos\theta\d\phi} \,.
\end{align}
%----

In Table~\ref{tab:Ai}, we present the angular coefficients $A_i$ for the $gg\to \ell^+\ell^-h$ process, considering the SM and new physics scenarios $\alpha=\pi/4$ and $-\pi/4$.
Two comments are in order. First, we observe sub-leading strong phase contributions from the one-loop calculation to the coefficients $A_{5,6,7}$, as evident from the SM scenario. Second, CP-violation effects are also depleted in the same coefficients as seen for the  $\alpha=\pi/4$ and $-\pi/4$ scenarios. Notably, the only statistically significant CP-phase $\alpha$ for the spin density 
parametrization arises in the coefficient~$A_0$.

%\bibliography{references} 
\bibliography{paper.bbl}

\end{document}